\documentclass[review]{elsarticle}
\usepackage{comment}
\usepackage{amsmath,amssymb,amsfonts}
\usepackage{graphicx}
\usepackage{subcaption}
\usepackage{textcomp}
\usepackage[ruled,vlined]{algorithm2e}
\usepackage{longtable}
\usepackage{lineno,hyperref}
\usepackage{graphicx}
\usepackage{subcaption}
\usepackage{amsmath}
\usepackage{longtable}
\usepackage{multirow}

\makeatletter
\def\ps@pprintTitle{%
  \let\@oddhead\@empty
  \let\@evenhead\@empty
  \let\@oddfoot\@empty
  \let\@evenfoot\@empty
}
\makeatother

\begin{document}

\begin{frontmatter}

\title{Poisoning Behavioral-based Worker Selection in Mobile Crowdsensing using Generative Adversarial Networks}

\author[label2,label3]{Ruba Nasser}%% or include affiliations in footnotes:
\author[label1]{Ahmed Alagha}
\author[label2,label3]{Shakti Singh}
\author[label2,label3]{Rabeb Mizouni \corref{cor1}}
\cortext[cor1]{I am the corresponding author}
\ead{rabeb.mizouni@ku.ac.ae}
\author[label2,label3]{Hadi Otrok }
\author[label2,label1]{Jamal Bentahar}

\address[label2]{Computer Science Department, Khalifa University, Abu Dhabi}
\address[label3]{Center of Cyber-Physical Systems (C2PS), Khalifa University, Abu Dhabi}
\address[label1]{CIISE, Concordia University, Montreal, Canada}

\begin{abstract}

With the widespread adoption of Artificial intelligence (AI), AI-based tools and components are becoming omnipresent in today's solutions. However, these components and tools are posing a significant threat when it comes to adversarial attacks. Mobile Crowdsensing (MCS) is a sensing paradigm that leverages the collective participation of workers and their smart devices to collect data. One of the key challenges faced at the selection stage is ensuring task completion due to workers' varying behavior. AI has been utilized to tackle this challenge by building unique models for each worker to predict their behavior. However, the integration of AI into the system introduces vulnerabilities that can be exploited by malicious insiders to reduce the revenue obtained by victim workers. This work proposes an adversarial attack targeting behavioral-based selection models in MCS. The proposed attack leverages Generative Adversarial Networks (GANs) to generate poisoning points that can mislead the models during the training stage without being detected. This way, the potential damage introduced by GANs on worker selection in MCS can be anticipated. Simulation results using a real-life dataset show the effectiveness of the proposed attack in compromising the victim workers' model and evading detection by an outlier detector, compared to a benchmark. In addition, the impact of the attack on reducing the payment obtained by victim workers is evaluated.

\end{abstract}


\end{frontmatter}

%\linenumbers

\section{Introduction}
\label{sec:introduction}

Mobile Crowdsensing (MCS) is a sensing approach that utilizes the collective involvement of mobile workers and their smart devices to collect data. The management platform first receives sensing tasks from task requesters and subsequently performs worker selection so that the quality of service (QoS) is maximized. The workers collect the requested data and send it back to the platform, which then pays the workers back for their contributed data. Artificial Intelligence (AI) based solutions have been widely adopted to optimize the performance of MCS. A notable real-world example is Uber, which leverages deep learning methods to optimize the Quality of Service (QoS) by predicting workers' estimated arrival times \cite{uber2024}. Waze is another well-known MCS application that utilizes data contributed by workers to enhance the driving experience by using AI methods to predict traffic and crash locations and timings \cite{waze2024}, \cite{AI_MCS_survey}.

One of the key challenges faced in MCS systems is selecting workers who are more likely to complete the tasks assigned. While they may initially accept these tasks, their behavior can vary significantly, leading to potential cancellations influenced by various contextual factors, such as the day of the task and the weather conditions. AI techniques have shown great potential in addressing this challenge. Using historical task data, supervised learning methods can be adopted to build behavioral models for each worker. These models can then be leveraged to predict the willingness of the workers to perform the task \cite{hussein2024crowd}. Despite the effectiveness of AI in optimizing MCS worker selection, their adoption introduces vulnerabilities that can be exploited by insider adversaries during the training phase \cite{AMLpoisoning}, \cite{jnca1}. Insider adversaries are individuals who have access to the management platform and exploit their trusted positions to alter the training process of the AI models by injecting malicious data \cite{jnca2}. This causes the models to behave incorrectly when deployed, ultimately undermining the system's fairness.

This work proposes an adversarial attack aimed at compromising behavior-based worker selection in MCS systems. The proposed attack bridges the theoretical advancements in adversarial machine learning by providing a novel application in practical settings like MCS worker selection and further highlights its real-world impact on the workers. In addition, one of the primary motivations for developing this attack model is to explore the risks presented by insider adversaries who have trusted access to the MCS platform and leverage it to manipulate the training process of the AI models. The proposed attack also highlights critical concerns about the reliability of AI-based selection systems due to their inherent vulnerabilities to malicious exploitation. These concerns are not merely theoretical; real-world cases have demonstrated the consequences of untrustworthy AI decisions. For example, Amazon had to discontinue its AI-based selection algorithm after discovering that it favored male over female candidates \cite{IBM_AI_Bias}. 

To this end, we propose a novel attack on behavioral-based MCS where malicious insiders leverage Generative Adversarial Networks (GANs) to generate poisoning points that can degrade the performance of victim workers' models by identifying vulnerable regions in the feature space of their data. GANs are powerful tools for data generation due to their unique adversarial training mechanism. A typical GAN architecture comprises two neural networks: the generator and the discriminator, trained simultaneously with opposing objectives in a minimax framework. The generator aims to produce synthetic data that closely resembles real data, while the discriminator learns to differentiate between genuine and generated samples \cite{gan_motiv}. Numerous GAN variants have been developed by incorporating additional components, and by modifying the loss functions to guide the generation process toward specific features or desired characteristics. In this study, we extrapolate on the method proposed in \cite{PganPaper} and build upon its framework to tailor the Poisoning GAN (PGAN) approach specifically for insider attacks on behavioral-based selection in MCS. The PGAN utilized in this study consists of a generator that tries to minimize the losses of both a discriminator and the victim worker's behavioral model during training \cite{PganPaper}. The insider then utilizes the trained PGAN to generate poisoning points and replaces a portion of the victim workers' data with the generated data. Subsequently, the platform uses the poisoned models to perform the selection, resulting in fewer task assignments to victim workers and a significant reduction in their overall revenue. While the attack can degrade the performance of the targeted models, it also considers the detectability constraints by generating poisoning points that can bypass outlier detectors. In addition, it does not compromise the QoS achieved for the tasks. Using a GAN-based approach makes the attack effective in terms of stealthiness, as it regulates the
generation of attack samples in a way that makes them unnoticeable. This allows the generated poisoning points to bypass anomaly detection models that can be adopted in the MCS system before the training stage. Overall, the main contributions of this work are summarized below:

\begin{itemize}
    \item A novel adversarial attack on behavioral-based MCS worker selection is proposed. In this attack, insider adversaries use PGANs to generate poisoning data to compromise victim workers' behavioral models during the training phase.
    \item Propose a targeted attack on MCS that specifically aims to lower the overall revenue of victim workers by decreasing the number of tasks they get assigned by the platform.
    \item Propose a novel framework that demonstrates how PGANs can be deployed at the worker selection stage to poison victim workers' models in MCS systems. The framework can be adopted to assess the efficacy of adversarial attacks against a resilient selection system that incorporates outlier detection and a QoS-based selection mechanism.
\end{itemize}

This work demonstrates how GAN-based poisoning attacks can be effectively adapted and leveraged against behavioral-based worker selection models in MCS systems, a scenario that, to the best of our knowledge, has not been previously explored.  Furthermore, by integrating and extrapolating PGAN in the MCS pipeline, this work reveals unique vulnerabilities in real-world worker selection systems, which have received limited attention in adversarial machine learning literature. 
Few studies investigated adversarial attacks on AI-based MCS, focusing mainly on the models adopted to improve the system's performance at the data aggregation stage and enhance its security \cite{AML_fake_task},\cite{AMLpoisongAttackDataAgreg}. To the best of our knowledge, this is the first work that proposes an adversarial attack specifically targeting behavioral-based selection models in MCS. Simulation results using a real-life dataset show the effectiveness of the proposed attack in compromising the victim workers' models and evading detection by an outlier detector, compared to a benchmark.

\section{Background and Related Work}
This section introduces GANs and discusses the related literature encompassing works on behavioral-based worker selection in MCS and adversarial attacks on AI-based MCS and IoT systems.
\subsection{ Background: Generative Adversarial Networks}
\label{sec:background}

GANs are a class of deep learning frameworks designed to generate new data samples that resemble a given training dataset. The typical GAN model consists of two neural networks: a generator $G$ and a discriminator $D$, trained in a competitive setting. The generator produces synthetic data samples, while the discriminator tries to distinguish between genuine and synthetic data. The training process continues until the generator produces samples that are indistinguishable from real data, effectively capturing the underlying data distribution.

The interaction between $G$ and $D$ can be modeled as a 2-player min-max game as shown in \eqref{eq:ganminmax}, where $V_{GAN}(D,G)$ is the objective function for $D$ that also depends on $G$. As defined in \eqref{eq:V}, $V_{GAN}(D,G)$ is the sum of the expected log-likelihood that the discriminator correctly identifies real and generated data, where $x$ is sampled from the real data distribution $p_x(x)$ and $z$ is sampled from a prior normal distribution $p_z(z)$ \cite{goodfellow2014generative}.
\begin{equation}
    \underset{G}{\min} \ \underset{D}{\max} \ V_{GAN}(D, G) 
    \label{eq:ganminmax}
\end{equation}

\begin{equation}
    V_{GAN}(D, G) = \mathbb{E}_{x \sim p_x(x)} [\log D(x)] + \mathbb{E}_{z \sim p_z(z)} [\log (1 - D(G(z)))]
    \label{eq:V}
\end{equation}

In GANs, the generator uses random noise as input, which can limit the model's performance and the quality of the generated data. To overcome this limitation, researchers have introduced auxiliary information to the input noise, enabling the generator to produce higher-quality data. Conditional GAN (CGAN) is one of the widely used types of GANs, and it uses class labels as input to both the generator and discriminator. Similar to GANs, the min-max game can be formulated as shown in \eqref{eq:ganminmaxcgan}, where the objective function for CGAN $V_{CGAN}(D,G)$ is defined in \eqref{eq:Vcgan}. The main difference between GAN and CGAN, is that in CGAN, $x$ is sampled from the real data distribution conditioned on class c $p_x(x|c)$ and $z$ is sampled from a prior normal distribution conditioned on the same class $p_z(z|c)$ \cite{gan_attacks_survey}. 

\begin{equation}
    \underset{G}{\min} \ \underset{D}{\max} \ V_{CGAN}(D, G) 
    \label{eq:ganminmaxcgan}
\end{equation}

\begin{equation}
    V_{CGAN}(D, G) = \mathbb{E}_{x \sim p_x(x|c)} [\log D(x|c)] + \mathbb{E}_{z \sim p_z(z|c)} [\log (1 - D(G(z|c)))]
    \label{eq:Vcgan}
\end{equation}

\subsection{GAN-based adversarial attacks}

The unique capabilities of GANs can be leveraged to manipulate and deceive machine learning models at various stages, including the pre-deployment or the post-deployment stage of the targeted models. In the pre-deployment stage, GANs are used to poison the training datasets by generating fake samples and injecting them into the datasets with flipped labels. Such techniques have been widely adopted in the computer vision domains \cite{LFgan1} and in federated learning systems \cite{LFgan2}, \cite{LFgan6}, \cite{LFgan7}. 
On the other hand, in the post-deployment stage, GANs can be utilized in multiple ways. The first is adversarial example generation, where the generator creates adversarial samples specifically designed to fool the trained classifier \cite{6}, \cite{109}, \cite{123}, \cite{154}. The other approach includes perturbation addition, where the generator creates subtle noise that, when added to the real data sample, results in increased classification error \cite{36}, \cite{46}. These methods are predominantly explored within the computer vision domain, where minimal alterations to test images often go unnoticed by human observers, yet can drastically impact model performance.

\subsection{Behavioral-based Worker Selection in MCS}
Several studies in the literature proposed AI techniques to perform behavioral-based worker selection in MCS systems. In  \cite{abououf2021machine}, unique models were trained for each worker to predict their willingness to perform the tasks. The models were trained using task-related data, such as the task start time, and worker-related data, such as the number of tasks completed per day. 
In \cite{gendy2020green}, an auctioning technique that adopts AI to predict workers' ability to complete sensing tasks was proposed. A Long Short-Term Memory (LSTM) model was leveraged to predict the battery levels and internet connectivity status of workers' devices throughout the sensing period. In \cite{nasser2022biometrics}, a biometrics-based selection framework was proposed, where a unique model for each MCS worker is built based on their unique interaction patterns with the smartphone’s touching
screen. By leveraging machine learning techniques, these
behavioral traits were used in order to detect impersonators in the system. 

Moreover, in \cite{zhu2020deep}, \cite{mcs_mob_pred}, and \cite{nasser2023machine}, historical mobility traces were used to predict workers' future location to improve worker selection in MCS. In \cite{zhu2020deep},
a deep learning-based approach was adopted to predict the future location values; then a greedy algorithm was used to perform the worker selection, such that the sensing coverage is maximized. In \cite{mcs_mob_pred}, deep learning was also used to perform the location prediction, and a weighted utility-based worker selection algorithm was proposed to perform the worker selection. Finally, in \cite{nasser2023machine}, a machine learning-based approach was proposed to predict workers’ future locations, which were then subsequently utilized in a continuous worker selection process based on a genetic algorithm.

Overall, the main advantages of the works discussed in this section include the use of AI-based methods that leverage historical worker data to predict their future behavior, thus optimizing the performance of MCS worker selection.  However, these methods are vulnerable to insider threats who can manipulate training data. This introduces risks that could significantly diminish the performance and reliability of the worker selection process.

\subsection{Adversarial Attacks on AI-based MCS and IoT Systems}

Adversarial machine learning is the study of how machine learning models can be manipulated by carefully crafted input. These inputs are intentionally designed by malicious adversaries to exploit the vulnerabilities of learning algorithms while causing models to make incorrect predictions. Adversarial attacks can be classified based on the phase of the machine learning pipeline in which they occur. In poisoning attacks, the adversary manipulates the training data to corrupt the learning process and degrade model performance. In contrast, evasion attacks involve crafting inputs that deceive a trained model during deployment without altering the training data \cite{AMLBackground}. 

Several studies explored insider attacks on AI models adopted in Internet of Things (IoT) systems, either before or after the model deployment. For instance, the attack proposed by \cite{smarthome_dos_aml} targeted ML-based intrusion detection systems in smart home networks. The authors proposed an approach to generate adversarial attack samples targeting the model after deployment, with the aim of misclassifying malicious network packets as normal. The techniques used to generate the adversarial samples include the Fast Gradient Sign Method (FGSM) and Jacobian Saliency Map Attack methods (JSMA). Another study, \cite{AML_malware}, proposed an adversarial attack on a machine learning-based malware detection system. The attack specifically targeted the model after deployment with the goal of successfully delivering the malware to smart devices. Techniques such as FGSM, JSMA, Carlini and Wagner (C\&W), and DeepFool were leveraged. 

Moreover, in \cite{AML_covid}, an insider attack on AI models deployed for medical diagnostic applications was proposed. The attack was conducted during the testing phase, and methods such as FGSM, DeepFool, C\&W, and JSMA were employed. The main goal of the attack is to increase the classification error of identifying Covid-19 cases. Finally, the attack proposed in  \cite{aml_enviornmentmonitoring} targeted models during the training phase, focusing on air quality monitoring applications. The utilized attack method is label-flipping, where the main goal is to degrade the performance of the ML model deployed to predict the impact of certain chemicals on the overall air quality.

A limited number of studies have proposed adversarial attacks on AI-based MCS systems. These works either aim to bypass security defense measures or degrade the quality of sensing achieved. In \cite{AMLpoisongAttackDataAgreg}, an unsupervised learning approach to generate poisoning data that degrades the performance of human activity recognition classifiers was proposed. The attack targets the models during the training phase and uses Self-Organizing Maps (SOM) to generate the poisoning data \cite{AMLpoisongAttackDataAgreg}. Furthermore, in \cite{AML_fake_task}, GANs were used to generate fake tasks that can successfully bypass machine learning-based fake task detection models. These models were trained to classify tasks into real and fake based on certain features such as task duration,  battery requirement, and start time. Since the features that characterize the fake tasks follow a certain distribution, the machine learning models can successfully identify them. However, the authors argue that adversaries could utilize CGANs to generate another type of fake tasks that are similar to real tasks, making them undetectable by the machine learning models and potentially overwhelming workers' devices.

The works discussed in this section emphasize the significant risks posed by insider threats within IoT and MCS environments. They proposed various attacks on AI models in various domains, including smart home networks, malware detection, medical diagnostics, environmental monitoring, and security. However, none of these studies specifically address the vulnerabilities related to insider adversaries targeting MCS worker selection and the subsequent implications for the overall revenue generated by workers from MCS tasks.

Table \ref{tab:LR} compares the works proposing adversarial attacks on machine learning models in MCS and IoT systems. As illustrated in the table, none of the existing works focused on developing an attack targeting MCS behavioral prediction models. This paper proposes an insider attack targeting victim workers to lower their revenue. The proposed attack uses PGANs, where the generator tries to increase the error of both the discriminator and the victim workers' behavioral models. Consequently, the trained generator can be used to craft poisoning data that are then injected by the insider into the victims' training datasets without being detected.

\begin{table}[h]
\caption{Summary of adversarial attacks on machine learning models in MCS and IoT systems}
\label{tab:LR} 
\begin{tabular}{|p{1.75 cm}|p{3cm}|l|p{3.5cm}|}
\hline
\textbf{Reference} & \textbf{Application} & \textbf{Phase} & \textbf{Method} \\ \hline
\cite{AML_fake_task} & MCS fake task detection & Post-deployment & CGAN \\ \hline
\cite{AMLpoisongAttackDataAgreg} & Activity recognition & Pre-deployment & SOM \\ \hline
\cite{smarthome_dos_aml}  & Intrusion detection & Post-deployment & FGSM and JSMA \\ \hline
\cite{AML_malware} & Malware detection & Post-deployment & FGSM, JSMA, C\&W, DeepFool \\ \hline
\cite{AML_covid} & Medical diagnosis & Post-deployment & FGSM, JSMA, C\&W, DeepFool \\ \hline
\cite{aml_enviornmentmonitoring} & Air quality monitoring & Pre-deployment & Label flipping \\ \hline
Proposed work & Behavioral-based worker selection & Pre-deployment & PGAN \\ \hline
\end{tabular}
\end{table}

\section{MCS Insider Attack Overview} 
One of the main reasons AI is adopted at the selection stage is to understand the behavior of the workers, which is done by collecting workers' data from previous tasks and training models to predict their willingness to perform future tasks. However, in this paper, we argue that malicious insiders can generate poisoning data to bias the worker selection using GANs. By replacing a portion of the historical data of the workers with the generated data, insiders can degrade the performance of victim workers' behavioral models by introducing misclassifications specifically related to the target class, without being detected. An overview of the proposed insider attack in the MCS platform is illustrated in Figure \ref{fig:overview}. Overall, the system includes the following modules:
\begin{figure} [t]

    \centering
    \includegraphics[width=1.2\linewidth]{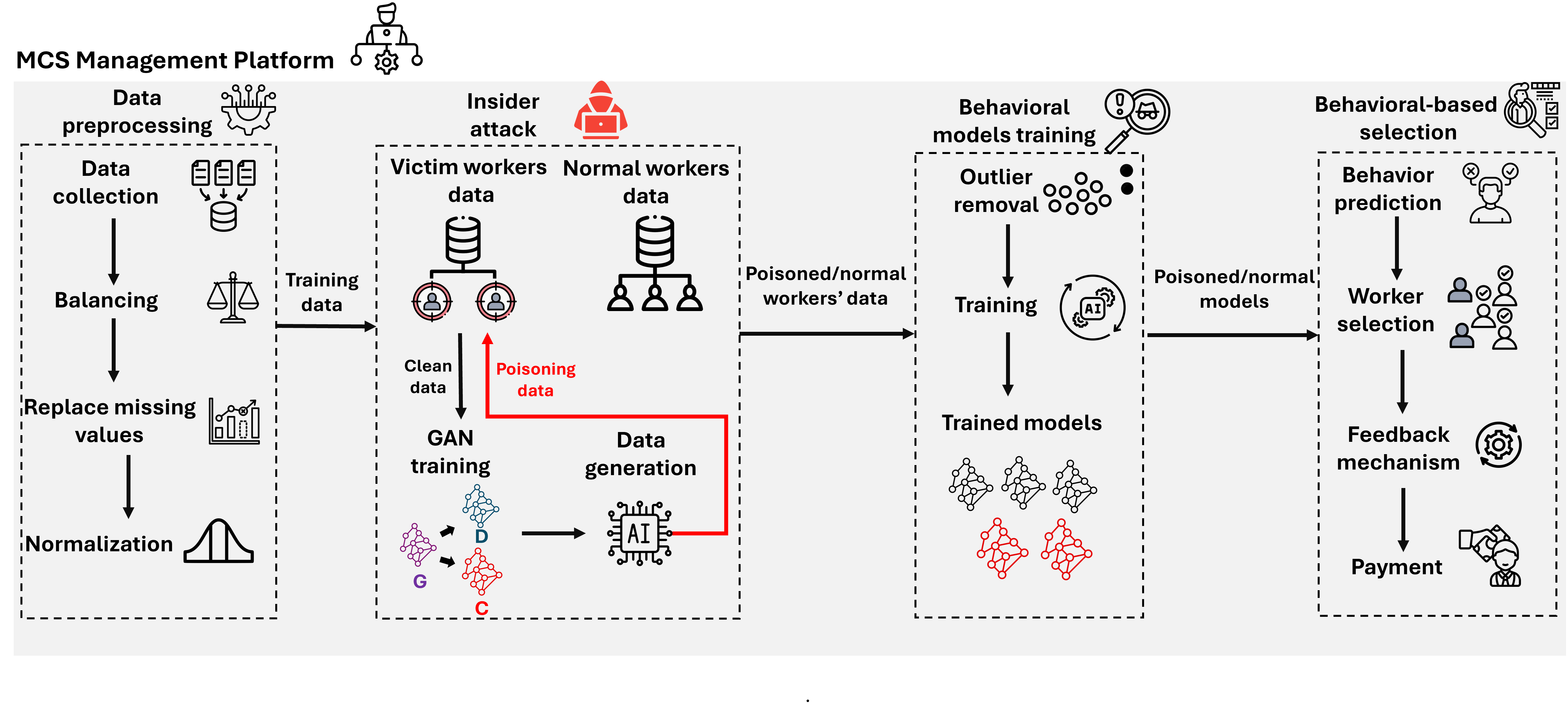}
    \caption{Overview of the proposed attack in the MCS platform, including 4 different modules: Data preprocessing, insider attack, behavioral model training, and behavioral-based selection module. }
    \label{fig:overview}
\end{figure}
\begin{itemize}

    \item \textbf{Data preprocessing:} This module is responsible for collecting data and preparing it for training. Initially, contextual data obtained from historical tasks is collected and stored in the platform. This data provides insights into various factors influencing task completion by the workers, and can be further categorized into worker-related features and task-related features. Each training point is labeled based on whether or not the tasks were canceled by the workers. Then, the training data undergoes preprocessing steps, which include data balancing using Synthetic Minority Over-sampling Technique (SMOTE), replacing missing values using Principal Component Analysis (PCA), and normalization. These steps are essential to ensure that the dataset is ready for training and the models yield optimal performance \cite{abououf2021machine}.
    
    \item \textbf{Insider attack:} In this module, an insider generates and injects poisoning points into the victim workers' training data to reduce their chances of being selected by the platform. First, a PGAN model is trained to generate poisoning points that are hard to detect by an outlier detector for each victim worker. The dataset used to train the PGAN model includes worker-related and task-related features, which provide insights into the workers' past performance and the context that can influence the workers' participation. The insider then uses the trained PGAN models to generate poisoning data and injects them into the victim workers' training datasets before their behavioral models are trained.

    \item \textbf{Behavioral model training:} This module is responsible for training workers' behavioral models, taking into consideration that malicious adversaries could inject poisoning data to skew the models' learning. Therefore, to ensure that the training data remains reliable and unbiased, the platform employs an Autoencoder-based outlier detection model to identify and remove anomalies from the workers' training datasets. Then, a deep learning model is built and trained for each worker to predict their behavior.  %However, malicious adversaries can still inject stealthy poisoning points that can bypass the outlier detectors and poison victim workers' models successfully. 

    \item \textbf{Behavioral-based selection:} This module uses the previously trained models to predict the workers' likelihood of canceling the task. Subsequently, a selection algorithm is executed to select a subset of workers such that the QoS of the task is maximized. After the task is completed, a feedback mechanism is deployed to update the selection metrics used based on the workers' performance. The feedback mechanism adopted also evaluates the QoS achieved by the selected workers to ensure it meets a satisfactory value for the task publisher. Moreover, the platform evaluates the selected workers' contributions and performs the payment accordingly. %As a result of the insider attack, victim workers receive lower payments due to their reduced likelihood of selection.

\end{itemize}

\section{The Proposed Poisoning Attack}

This section presents the proposed attack model, demonstrating the vulnerability of the behavioral models in MCS systems. It also describes the features used, the preprocessing steps performed, and the training process of the PGAN model, as described in the following subsections.

\subsection{Threat Model}

Let $ W = \{w_1, w_2, \dots, w_k\} $ be a set of workers in the MCS system, where each worker $ i$ has a training dataset $ D_{i} = \{(x_j, y_j) \mid x_j \in \mathbb{R}^{n}, \, y_j \in \{0, 1\}, \, j = 1, \dots, m \} $ that captures their behavior in terms of the willingness to complete the tasks assigned. The dataset consists of n-dimensional feature vectors $x_j$ and their corresponding class labels $y_j$,  where $y_j = 1$ indicates that the task was canceled by the worker and $y_j = 0$ indicates otherwise. The platform trains a behavioral model $M_{i}$ for each worker, using their dataset $D_{i}$. In the proposed attack, an insider adversary seeks to compromise the models of a set of victim workers $V$, where $ V \subset W $, by injecting poisoning points $\mathcal {P}_{i}$
to the training dataset $\mathcal{D}_{i}$. The objective is 
to classify data points belonging to class $y_j=0$ as $y_j=1$, i.e. predict that a worker will cancel the task when they actually are more likely to accept it.

\subsection{Features Description and Data Preprocessing}

For each victim worker $i$, the insider trains a PGAN model using the original dataset $\mathcal{D}_{i}$. The features used in each dataset $\mathcal{D}_{i}$ can be classified into worker-related features and task-related features, both of which can be used for effectively predicting the likelihood of task cancellation by the worker $i$. The former provides insights into the workers' past performance and ability to complete the tasks, such as the cumulative rating and assigned workload on the day of the task. On the other hand, the latter includes features related to the tasks assigned and provides context that can influence the likelihood of successful completion, such as the weather conditions and the starting time of the task. In this work, the dataset, including the behavioral data of all workers utilized for training the PGAN models, is explained in Section \ref{sec:results}.

Every dataset $D_{i}$ stored in the platform undergoes the following preprocessing steps. Firstly, PCA is employed to address the issue of missing values in the dataset. Initially, the mean is used to replace the missing values, and then PCA is applied to transform the data to a lower dimensional space. Missing values are estimated after the data is transformed back to the original space based on the relationship identified between the variables when PCA was applied. Secondly, SMOTE is used to balance the dataset. This technique generates synthetic samples for the minority class through interpolation between existing instances and their nearest neighbors. 
Finally, Min-Max normalization is applied to scale all feature values within the range $[0, 1]$ \cite{abououf2021machine}.

\subsection{Poisoning Points Generation using GANs}

The proposed attack targets a deep learning behavioral model $M_{i}$ that outputs the probability of task cancellation $PC_{i}$ by a victim worker $ i \in V $. The model takes a feature vector ${x} $ as input and returns the probability of cancellation, as defined in \eqref{eq:prob}.

\begin{equation}
   M_{i}({x}) = PC_{i}
   \label{eq:prob}
\end{equation}
The predicted class label $\hat{y}_{i}$ is determined by applying a threshold to the model's output, as given by \eqref{eq:yhat}.

\begin{equation}
   \hat{y}_{i} = 
\begin{cases}
1 & , \text{if } PC_{i} \geq 0.5, \\
0 & , \text{otherwise } %PC_{i} < 0.5,
\end{cases}
\label{eq:yhat}
\end{equation}

Before $M_{i}$ is trained in the MCS platform, an insider uses $\mathcal{D}_{i}$ to train a PGAN model and leverages it to generate a set of poisoning points $\mathcal{P}_{i}$. The main goal of the attack is to degrade the performance of $M_{i}$ when trained using the poisoned dataset by increasing the error rate of misclassifying points from class 0 into the target class $y_t=1$. The attacker injects the points $\mathcal{P}_{i}$ into the training dataset $\mathcal{D}_{i}$ by replacing a portion of the data points belonging to target class $y_t = 1$ with $\mathcal{P}_{i}$.

%In addition, the attack increases the predicted probabilities of task cancellation for the input features belonging originally to class 0. For some instances, this causes the probabilities to go beyond the threshold value, resulting in a flip in the label from $\hat{y}_{i}=0$ to $\hat{y}_{i}=1$, for others only the probability value increases without flipping the label.

\begin{figure}
    \centering
    \includegraphics[width=0.70\linewidth]{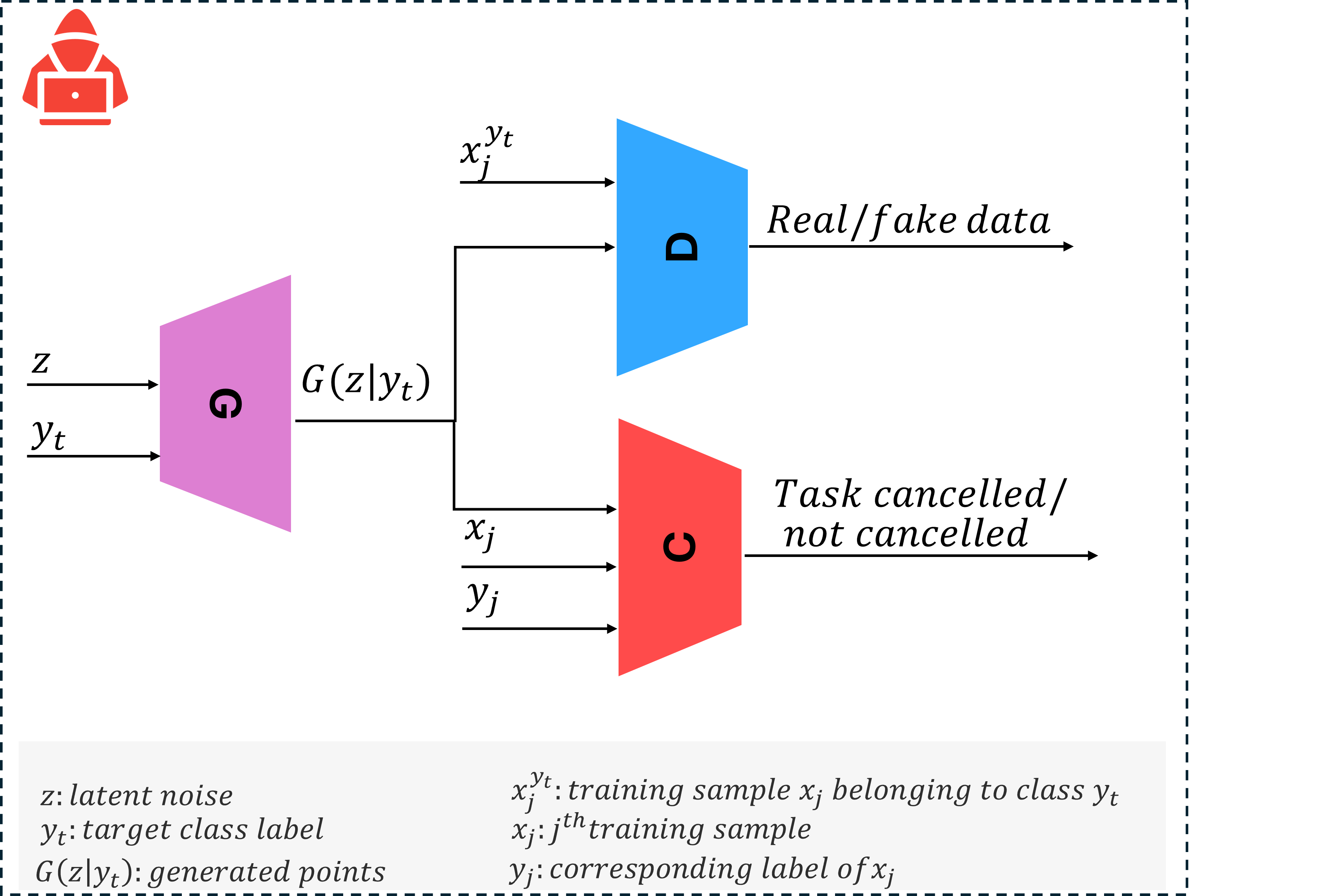}
    \caption{PGAN model representation}
    \label{fig:pgan}
\end{figure}

As discussed in Section \ref{sec:background}, GANs and CGANs can be used to generate realistic data. This functionality can be exploited for malicious purposes by adding a third component to the GAN model, enabling the generation of adversarial data.
The PGAN model used in this work comprises 3 main components: a generator $G_{i}$, a discriminator $D_{i}$, and a classifier $C_{i}$, as shown in Figure \ref{fig:pgan}. The PGAN model is trained adversarially, where $G_{i}$ competes against both $C_{i}$ and $D_{i}$. During training, $G_{i}$ aims to create points that increase the losses of both $C_{i}$ and $D_{i}$. As a result, $D_{i}$ exhibits a higher error at distinguishing between original and generated points, and $C_{i}$'s error in predicting workers' behavior increases. On the other hand, $D_{i}$ and $C_{i}$ aim to minimize their respective losses, despite the $G_{i}$'s attempts to disrupt their performance. 

The interaction between $G_{i}$, $D_{i}$, and $C_{i}$ can be modeled as min-max game as shown in \eqref{eq:pganminmax}, where $V_{PGAN}(D_{i},G_{i})$ is the objective function for $D_{i}$ and $W_{PGAN}(C_{i},G_{i})$ is the objective function for $C_{i}$. 

\begin{equation}
    \underset{G_{i}}{\min} \ \underset{D_{i},C_{i}}{\max} \ \alpha V_{PGAN}(D_{i},G_{i}) + (1- \alpha) W_{PGAN}(C_{i},G_{i}) 
    \label{eq:pganminmax}
\end{equation}

According to the formula in \eqref{eq:pganminmax}, $D_{i}$ tries to maximize its objective function $V_{PGAN}(D_{i},G_{i})$ and $C_{i}$ tries to maximize its objective function $W_{PGAN}(C_{i},G_{i})$. Note that both objective functions depend on $G_{i}$. Moreover, $G_{i}$ tries to minimize the weighted sum of the objective functions $V_{PGAN}(D_{i},G_{i})$ and $W_{PGAN}(C_{i},G_{i})$, where $\alpha$ is the weighting factor that determines the relative contribution of both. By minimizing the combined objectives, $G_{i}$ generates points that increase the error of $D_{i}$, and $C_{i}$, making $D_{i}$ less effective at distinguishing genuine and generated points, while degrading $C_{i}$'s ability to predict the worker behavior correctly.

The parameter $\alpha$ plays a crucial role on the attack effectiveness and in shaping the distribution of the generated points. At $\alpha = 1$, the PGAN model behaves as a CGAN and its generator generates points that are highly similar to the target class $y_t$. However, this comes at the cost of reduced attack effectiveness as the generated data points fail to degrade the behavioral model's performance. On the other hand, when $\alpha=0$, the PGAN's generator generates points that are further from the target class $y_t$ and exhibit greater similarity to the other class, thus making the attack more effective. By carefully selecting the optimal value of $\alpha$, which lies between 0 and 1, the attacker can train a PGAN model to generate subtle poisoning points that compromise the worker's behavioral model. Moreover, the attacker can identify the $\alpha$ value, which effectively exploits the decision boundary of the behavioral model, enabling the poisoning points to successfully execute the targeted attack by misclassifying data points from class 0 as 1.   
%The objective is to classify data points belonging to class $y_j=0$ as $y_j=1$, i.e. predict that a worker will cancel the task when they actually are more likely to accept it.

As defined in \eqref{eq:VPGAN}, $V_{PGAN}(D_{i},G_{i})$ is the sum of the expected log-likelihood that the discriminator correctly identifies real and generated data, where $x$ is sampled from the real data distribution $p_x(x|y_t)$ conditioned on the target class $y_t$ and $z$ is sampled from a prior normal distribution $p_z(z|y_t)$ conditioned on the same class $y_t$. The first term represents the log-likelihood that the discriminator correctly classifies real points, where $D_{i}(x|y_t)$ is the probability that the $x$ is real. On the other hand, the second term represents the log-likelihood that the discriminator identifies the generated data $G_{i}(z|y_t)$ as fake, where $1 - D(G_{i}(z|y_t))$ is the probability that the generated data are fake. 

\begin{equation}
\begin{split}
    V_{PGAN}(D_{i}, G_{i}) = & \, \mathbb{E}_{x \sim p_\text{data}(x|y_{t})} [\log D_{i}(x|y_t)] \\
    & + \mathbb{E}_{z \sim p_z(z|y_t)} [\log (1 - D(G_{i}(z|y_t)))].
\end{split}
\label{eq:VPGAN}
\end{equation}
In addition, $W_{PGAN}(C_{i},G_{i})$, is expressed in \eqref{eq:WPGAN}, where $L_{C_i}$ is the loss function used to train the classifier $C_{i}$ and  $\lambda \in [0,1]$ is a weighting factor that balances the classifier’s performance between classifying the poisoning points and the original training samples. During PGAN training, the classifier $C_{i}$ is exposed to both poisoning points and original training points. The term $\mathbb{E}_{z \sim p_z(z|y_t)} [L_{C_{i}}(G_{i}(z|y_t))] $ evaluates the expected loss of the classifier on the poisoning points generated by the generator. In addition, the term $\mathbb{E}_{x \sim p_x(x)} [L_{C_{i}} (x)]$ represents the expected loss of the classifier on the original points. The best value of $\lambda$ should be chosen such that the classifier's goal is to perform well on both original and poisoning points.
\begin{equation}
\begin{split}
    W_{PGAN}(C_{i}, G_{i}) = & -( \lambda  \mathbb{E}_{z \sim p_z(z|y_t)} [L_{C_{i}}(G_{i}(z|y_t))] \\ & + (1-\lambda) \mathbb{E}_{x \sim p_x(x)} [L_{C_{i}} (x)])
    \label{eq:WPGAN}
\end{split}
\end{equation}

%This attack assumes the insiders have knowledge of the architecture of the model $M_{i}$, which is a realistic assumption given their position. 
% During GAN training, the generator plays a min-max zero-sum game with the discriminator and classifier, aiming to generate poisoning points that increase the classifier’s error while remaining similar to the original data. On the other hand, the classifier seeks to minimize its error on a poisoned dataset, and the discriminator aims to differentiate between the generated and original data. Once the training is completed, the insider uses $G_{i}$ to generate $\mathcal{P}_{i}$ and injects them into the dataset $\mathcal{D}_{i}$.

%The generator network includes an embedding layer that converts the discrete target class label \( y_t \) into a continuous vector representation, enabling the generator to incorporate class-specific information. This representation is concatenated with the random noise \( z \) to form the conditional input, which is then passed through a series of fully connected layers to produce the final generated samples \( \tilde{x} \).
Algorithm \ref{alg:pgan_training} shows the pseudocode of the training process of the PGAN model, where the models $G_{i}$, $D_{i}$, and $C_{i}$ are iteratively trained. In each iteration, a mini-batch of random noise samples $z$ is sampled, which is then used by $G_{i}$ to generate the points $\tilde{x}$. The discriminator then uses the generated data $\tilde{x} $ and a mini-batch of training samples ${x_{y_t}}$, which belong to class $y_t$, and learns to differentiate between real and generated data by minimizing a loss function $L_{D_i}$. $L_{D_i}$ is formulated as shown in ~\eqref{eq:L_D}, where $y_f$ denotes the labels of the fake data generated by $G$, set to 0, $y_r$ denotes the labels of the real data, set to 1, and $L_{CE}$ represents the binary cross-entropy loss function.

\begin{algorithm}[h]
\caption{Training Process for PGAN}
\label{alg:pgan_training}
\KwIn{target class $y_t$, mini-batch size $m$, $\lambda$, $\alpha$, number of iterations $M$}
\KwOut{Trained generator $G_{i}$, discriminator $D_{i}$, and classifier $C_{i}$}

Build $G_{i}$, $D_{i}$, and $C_{i}$ models

\For{iteration \( j = 1, 2, \dots, M \)}{
    Sample a mini-batch of \( m \) random noise samples \(z\)

    Use \( G_{i} \) to generate points \( \tilde{x} \), where \( \tilde{x} = G_{i}(z|y_t) \)
    
    Select a mini-batch of \(m\) samples \({x}_{y_t} \) belonging to class \( y_t \) 
    
    Use \( D_{i} \) to make predictions on \( {x}_{y_t} \) and \(\tilde{x} \)

    Compute discriminator loss \( L_{D_i} \) using \eqref{eq:L_D}
    
    Update \( D_{i} \) by minimizing \( L_{D_i} \)
    
    Select a mini-batch of \(m\) training features \( x \) and labels \( y \)

    Use \( C_{i} \) to predict the behavior of worker \(i\) using \(\tilde{x}\) and  \(x\)
    
    Compute the classifier loss \( L_{C_i} \) using \eqref{eq:L_C}
    
    Update \( C_{i} \) by minimizing \( L_{C_i} \)
    
    Compute generator loss using \eqref{eq:L_G}
    
    Update \( G_{i} \) by minimizing \( L_G \)
}

\textbf{Return} trained \( G_{i} \), \( D_{i} \), and \( C_{i} \)

\end{algorithm}

\begin{equation}
L_{D_i} = L_{CE}(D_{i}(\tilde{x}),y_f)+ L_{CE}(D_{i}(x_{y_t}),y_r)  
\label{eq:L_D}
\end{equation}

Subsequently, the classifier uses a mini-batch of training features $x$, their corresponding class labels $y$, and the generated data $\tilde{x}$, to learn the behavior of the victim worker. This is achieved by minimizing a loss function $L_{C_i}$, which is formulated in \eqref{eq:L_C}.

\begin{equation}
L_{C_i} = \lambda L_{CE}(C_{i}(\tilde{x},y_t)+ (1-\lambda) L_{CE}(C_{i}(x),y)  
\label{eq:L_C}
\end{equation}

Finally, the generator model parameters are updated by minimizing the loss function $L_{G_i}$, defined in \eqref{eq:L_G}, where $\alpha \in [0,1]$ controls the shape of the distribution of the generated points \cite{PganPaper}.

\begin{equation}
L_{G_i} = \alpha L_{CE}(D_{i}(\tilde{x}),y_r)+ (1-\alpha) L_{CE}(C_{i}(\tilde{x}),(1-y_t))  
\label{eq:L_G}
\end{equation}

\subsection{Behavioral models training and outlier detection}

%This section provides a detailed explanation of the outlier detection technique employed; in addition, it describes the behavioral-based selection approach adopted.

% \subsection{Outlier removal}
After training a PGAN model for each victim worker, the insider utilizes the trained generator \( G_{i} \) to produce poisoning data \( \mathcal{P}_{i} \). This generated data is then injected into the training dataset \( D_{i} \) by replacing the original data points belonging to class \( y_t \) with \( \mathcal{P}_{i} \).
Before training the workers' behavioral models, the platform attempts to detect and remove outliers from each class in $D_{i}$ to prevent biased model decisions. 

To achieve this, an autoencoder was trained on the target class data, where an encoder learns a lower dimensional representation of the input, and then a decoder reconstructs the original data from the encoded vector. The target class data is then passed to the trained autoencoder model, and the reconstruction error for each data point is computed. Data points with high reconstruction errors are identified as outliers, based on a predefined threshold \cite{AEoutlierdetector}.

%To achieve this, a dimensionality reduction technique is applied to reduce the high dimensional training features $x$ into 2 dimensions. The dimensionality reduction technique used in this work is the t-Distributed Stochastic Neighbor Embedding (t-SNE) \cite{tsneOutlierRemoval}. t-SNE is a non-linear technique that maps high-dimensional data into a lower-dimensional space. It works by computing similarities between points in the original space and then arranging the points in the reduced space to preserve those similarities. Specifically, t-SNE minimizes the Kullback-Leibler (KL) divergence between the probability distributions in the original and reduced spaces, ensuring that points with similar relationships in the original space remain close in the 2D space \cite{van2008visualizing}.Once the data is reduced to 2 dimensions, the centroid of each class is evaluated. Subsequently, the Euclidean distance is calculated between each data point and its class centroid. Finally, a threshold is applied to these distances to identify and remove outliers. Points with distances exceeding the threshold are considered outliers and are removed from the dataset. However, the outlier detection process fails to identify the injected points as outliers by the attack carried out by insiders. As a result, victim workers' models are poisoned, leading to biased decisions and compromised performance.

After removing outliers, every dataset $D_{i} = \{(x_j, y_j) \mid x_j \in \mathbb{R}^{12}, \, y_j \in \{0, 1\}, \, j = 1, \dots, n_i \} $ is used to train a deep learning model $M_{i}$ to predict the behavior of the corresponding worker $w_{i}$. The training process involves minimizing a loss function $L_{i}$, which evaluates the classification error, given the predicted output $\hat{y}_j$ and the true output $y_j$. The loss function used in this work is Binary Cross Entropy (BCE), which is also shown in \eqref{eq:L_i}. As a result, the parameters of the model $M_{i}$ are adjusted so that the model can be used to make accurate predictions on new data.

\begin{equation}
    L_{i} = - \frac{1}{n_i} \sum_{j=1}^{n_i}  y_j \log \hat{y}_j + (1 - y_j) \log (1 - \hat{y}_j) 
    \label{eq:L_i}
\end{equation}

After training, each model is used at the selection stage to predict the task cancellation probability $pc_{i}$. The platform uses this probability along with other parameters to select a group of workers, as described in the following section.

\section{Behavioral-based selection}
\label{sec:behavioral selection}
Let $ W = \{w_1, w_2, \dots, w_n\}$ denote the set of workers in the MCS system, where \( i \) can be mathematically represented as a tuple: $i = (id_{i},r_{i}, {l}_{i}, M_{i})$.
Every worker is characterized by the following attributes: a unique identifier $id_{i}$,
a reputation score $ r_{i} \in [0, 1] $, the geographic location ${l}_{i}$, and the trained behavioral model $M_{i}$. 
The platform aims to select a subset of workers $g$, where $g \subset W$, to perform a given task $k$, such that the QoS is maximized. Each task $k$ is represented as a tuple $k=(id_k, t_k, td_k, l_k)$, where $id_k$ is the task id, $t_k$ is the task's starting time, $td_k$ is the task deadline, and $l_k$ is the task's location.

\subsection{The selection approach}
The main contribution of this work is to propose a novel framework that shows how existing behavioral-based worker selection can be compromised through adversarial attacks using GANs. Therefore, a greedy selection algorithm is employed to find the subset of selected workers $g$ based on their QoS values, as described in Algorithm \ref{alg:1}. The QoS score for each worker $QoS_{i}$ considers multiple factors, including latency $\tau_{i}$, reputation $r_{i}$, and completion confidence $conf_{i}$, as defined in \eqref{eq:qosi}.

\begin{equation} QoS_{i} = \tau_{i} \times r_{i} \times {conf}_{i} \label{eq:qosi} \end{equation}

The parameter $\tau_{i}$ represents how quickly the worker can arrive at the task location and has a decreasing value with the increase in the time required to reach the task. It can be evaluated as shown in \eqref{eq:taw}, where $tt_i^{k}$ is the worker's traveling time to the task estimated based on the current location $l_{i}$ and $td_k$ is the deadline by which task $k$ should be completed. 
\begin{equation}
    \tau_{i}=[ 1 - \max ( 0, \min ( \log_{td_k}(tt_i^{k}), 1 ) ) ]
    \label{eq:taw}
\end{equation}  
Additionally, $r_{i}$ denotes the reputation value, reflecting the workers' current and historical performance, as shown in \eqref{eq:r_i}.

\begin{equation}
    r_{i}=\gamma r_{i} + (1-\gamma) \Omega_{i}
    \label{eq:r_i}
\end{equation}

$r_{i}$ denotes the reputation value before the completion of the task, whereas $\Omega_{i}$ represents the worker's most recent performance and is calculated as the percentage of successfully completed tasks, as shown in \eqref{eq:r}.
The reputation score is updated by taking the weighted sum of these 2 parameters, where $\gamma$ is the weight value given to the old reputation value.

\begin{equation}
     \Omega_{i} = \frac{\text{Num of successfully completed tasks by worker $i$}}{\text{Num of assigned tasks to worker $i$}} 
    \label{eq:r}
\end{equation}

The third parameter used in the calculation of the QoS is $conf_{i}$, which represents the worker's confidence in completing the task. It can be evaluated as shown in \eqref{eq: confi}
\cite{abououf2021machine}. 

\begin{equation}
    conf_{i}= 1- pc_{i}
    \label{eq: confi}
\end{equation}

\begin{algorithm}[H]
\caption{Greedy Worker Selection}
\label{alg:greedy_selection}

\SetAlgoLined
\KwIn{Set of workers $W$, task \( k \), group size \( GroupSize \)}
\KwOut{Selected subset of workers \( S \) to perform task \( k \)}

Initialize \( g \gets \emptyset \) \tcp*{Start with an empty set of selected workers}

\For{each worker \( i \in W \)}{
    
    Evaluate worker latency \( \tau_{i} \) using Equation~\eqref{eq:taw}\;
    
    Calculate worker reputation \( r_{i} \) using Equation~\eqref{eq:r}\;
    
    Compute QoS for worker \( i \) using Equation~\eqref{eq:qosi}\;
}

Sort workers in \( W \) by their \( QoS_{i} \) values in descending order\;

\While{$|S| < GroupSize $ and \( W \) is not empty}{
    Select the worker \( i \) with the highest \( QoS_{i} \) from \( W \)\;
    
    Add \( i \) to \( g \)\;
    
    Remove \( i \) from \( W \)\;
}
\label{alg:1}
\end{algorithm}

\subsection{Feedback mechanism and payment evaluation}

After completing a task, the system updates the reputation values based on the workers' performance, as shown in \eqref{eq:r_i}.
In addition, the feedback mechanism also includes evaluating and monitoring the QoS of the selected group of workers ${QoS}_g$, to identify potential security breaches that may have occurred \cite{li2019QoSAnomaly}. The QoS metric is evaluated for a group of workers rather than for individual workers because MCS tasks are typically carried out collaboratively. Therefore, the service quality and reliability of the sensing outcome depend on the combined efforts and collective contributions of all members within the group. This step is critical to ensure the reliability of the system and support early detection of insider threats. 
Therefore, the value of $QoS_g$ can serve as feedback, triggering an alarm for potential manipulation in the platform. However, the proposed attack does not affect the QoS values, demonstrating its ability to bypass this additional layer of detection. %i'm not very convinced with this. Monitoring Qos will be a way to detect insiders who want to harm the system by degrading its value. This is not the insider attack I am proposing. QoS_g will always be high because the algorithm will always select the best workers. The intent of the attack is not to harm the QoS but to lower the payment of the workers. 
$QoS_g$ can be evaluated as illustrated in \eqref{eq:qosg}, where $r_g$, $\tau_g$, and $conf_g$ are the reputation, time score, and confidence of the selected group, respectively. 

\begin{equation}
    QoS_g=w_1  r_g + w_2  \tau_g+ w_3 conf_g
    \label{eq:qosg}
\end{equation}

$r_g$ and $conf_g$ can be evaluated as shown in \eqref{eq:r_g} and \eqref{eq:conf_g}, respectively,  by taking the minimum value of the reputation and confidence of the workers in the group.  

\begin{equation}
    r_g = \min_{i \in g} \{ r_{i} \}
    \label{eq:r_g}
\end{equation}

\begin{equation}
    conf_g = \min_{i \in g} \{ conf_{i} \}
    \label{eq:conf_g}
\end{equation}

In addition, the latency of the group is evaluated as shown in \eqref{eq:taw_g}, where $|g|$ is the group size. It is calculated by taking the product of the average latency and the exponential of its negative standard deviation. This approach ensures that the overall value decreases if there is more variability in the individual scores \cite{abououf2021machine}.

\begin{equation}
    \tau_g =  (\frac{1}{|g|} \times \sum_{i \in g} \tau_{i} )\times e^{-\sigma(\tau_{i})}
    \label{eq:taw_g}
\end{equation}

After the evaluation of $Qos_g$, workers are paid based on their contributions to the sensing task. The payment for each user is calculated as in \eqref{eq:payment}.

\begin{equation}
    \text{Payment} = \mu + \left( \frac{QoS_g - QoS_i}{QoS_g} \times \text{BP} \right)
    \label{eq:payment}    
\end{equation}

The payment value is determined by multiplying a base payment $BP$ by the worker's contribution to the task, which is calculated as the ratio of the difference between \( QoS_g \) and \( QoS_i \) to \( QoS_g \). This amount is then added to a fee $\mu$ that adjusts the price based on traffic conditions to encourage workers to participate \cite{nasser2023machine}.

%\subsection{Data visualization}
%This section presents a visualization of a worker's dataset after dimensionality reduction, highlighting the impact of varying the value of $\alpha$.
%Since the dataset $D_i$ consists of 12-dimensional features, a dimensionality reduction technique is required to visualize the data in two dimensions. The technique used in this work is t-Distributed Stochastic Neighbor Embedding (t-SNE). The dataset used to train the GAN model is the ride Austin dataset \cite{abououf2021machine}, which is further described in section \ref{}. 

\section{Results}
\label{sec:results}

This section presents the simulation results showing the effectiveness of the proposed poisoning attack. Firstly, the Ride Austin dataset was filtered by worker ID, and each subset was treated as an individual training dataset. Subsequently, 86 PGAN models were individually trained, one for each worker, to generate the poisoning samples, which were then injected into the corresponding worker’s training dataset. This is done to ensure that the robustness of the attack is validated across a diverse set of 86 workers, and to prove that it is generalizable across different behavioral patterns.

Four experiments were conducted, each designed to evaluate a distinct aspect. The first experiment examines the effect of $\alpha$ on the models' performance and on the distribution of the generated poisoning points. The second experiment evaluates the impact of varying poisoning percentages on the models' performance. The third experiment compares the proposed approach with two existing benchmarks by assessing the detectability of the attack by an outlier detector and evaluating the models' performance for varying poisoning percentages. Lastly, the fourth experiment analyzes how the attack affects worker selection. 

\subsection{Model architecture and training setup}

The PGAN model used in our experiments consists of three main components: a generator, a discriminator, and a classifier. The generator network comprises four fully connected layers. The first layer combines the input noise and conditional label information, mapping them to a 100-dimensional vector. The subsequent layers are dense fully connected layers with 784 and 1024 neurons, respectively. The final layer maps the output to 12 dimensions, matching the dataset's dimensionality. Each layer uses Leaky ReLU activations, and the final layer applies a sigmoid activation. 

The discriminator and classifier architectures are similar, each consisting of four fully connected layers. The first layer combines the input data and conditional label, mapping them to a 784-dimensional vector. The next two layers are dense layers with 1024 and 512 neurons, respectively. The final layer outputs a scalar value representing the probability that the input is real or fake. Each layer uses Leaky ReLU activations, and the final output is passed through a sigmoid activation to produce a probability between 0 and 1. 

The total number of trainable parameters in the proposed GAN-based model is approximately 4.29 million. This includes parameters from the Generator, Discriminator, and Classifier networks. These values demonstrate the complexity of the model, reflecting the number of learnable parameters that are optimized during the training process. Therefore, the model used is considered significantly less complex, compared to more advanced models, such as ResNet, which includes up to 11 million parameters. To train the PGAN models, an NVIDIA Tesla V100 GPU with memory of 32 GB was used. Moreover, the time taken to train a PGAN model for one worker is around 30 minutes.

Each PGAN model was trained for a total of 2000 epochs and with $\lambda = 0.8$. By setting the value of $\lambda$ to 0.8, the classifier places more importance on minimizing the loss obtained from the generated data points while still maintaining some importance on the performance of the classifier on real data points. So as the training progresses, it becomes better at classifying the poisoning points correctly, which in turn drives the generator to enhance its ability to generate better poisoning points that can mislead the classifier.

\subsection{Dataset}

The experimental results presented in this section utilize a real-life dataset to ensure that they effectively illustrate the impact of the attack on the models of workers exhibiting diverse behaviors in real-life scenarios.
The dataset used in this work is the Ride Austin dataset \cite{abououf2021machine}. It contains rides completed and canceled by a total of 86 workers over a period of 8 months in Austin, Texas, USA. This dataset also includes features that can be classified into worker-related or task-related features, as shown in Table \ref{tab:features}. The worker-related features include the number of assigned and completed tasks on the day of the task, the worker rating, and the car rating. The worker rating represents the cumulative evaluation of the workers by the task requestors based on their historical performance. Similarly, the car rating is the cumulative rating of the worker's vehicle over previous tasks. Furthermore, task-related features include the task starting time, requestor's rating, price adjustment fee, and features representing the weather conditions on the day of the task, such as the precipitation, maximum and minimum temperature, wind speed, and wind gust. The requestor's rating comprises the cumulative evaluation of the task requestor by the workers, whereas the price adjustment fee represents the additional amount of money added to the workers' payment in response to real-time conditions, such as traffic, to motivate them to perform the task.

\begingroup
\footnotesize

\begin{longtable}{|p{1.5cm}|p{2cm}|p{7.5cm}|}
\caption{Features used in the dataset} \label{tab:features} \\
\hline
\textbf{Category} & \textbf{Feature} & \textbf{Description} \\
\hline
\endfirsthead

\hline
\textbf{Category} & \textbf{Feature} & \textbf{Description} \\
\hline
\endhead

\hline
\endfoot

\hline
\endlastfoot

Worker-related
  & Number of assigned tasks & The total number of assigned tasks on the day of the task. \\
  \cline{2-3}
  & Number of completed tasks & The total number of completed tasks on the day of the task. \\
  \cline{2-3}
  & Worker rating & The cumulative evaluation of the workers by the task requestors. \\
  \cline{2-3}
  & Car rating & The cumulative evaluation of the worker's vehicle condition by the task requestors. \\
\hline

Task-related
  & Start time & The tasks' starting time. \\
  \cline{2-3}
  & Requestor rating & The cumulative evaluation of the task requestor by the workers. \\
  \cline{2-3}
  & Price adjustment fee & An additional amount added to the workers' payment in response to real-time conditions like traffic to encourage them to participate. \\
  \cline{2-3}
  & Precipitation & A measure of the amount of rain falling at the time of the task. \\
  \cline{2-3}
  & Maximum temperature & The maximum temperature on the day of the task. \\
  \cline{2-3}
  & Minimum temperature & The minimum temperature on the day of the task. \\
  \cline{2-3}
  & Wind speed & The speed of the wind at the time of the task. \\
  \cline{2-3}
  & Wind gust & A sudden brief increase in wind speed beyond the average speed at the time of the task. \\
\hline

\end{longtable}

\endgroup

\subsection{Performance evaluation metrics}

To evaluate the effectiveness of the proposed attack on the performance of the models, the following metrics were used: False Positive Rate (FPR), False Negative Rate (FNR), Accuracy, Precision, Recall, and F1 Score. FPR measures the proportion of accepted tasks incorrectly predicted as canceled, while FNR is the proportion of canceled tasks incorrectly predicted as accepted (not canceled). %where the false positives represent the number of tasks incorrectly predicted as canceled, and the true negatives represent the number of tasks correctly predicted as not canceled. 
%where the false negatives represent the number of tasks incorrectly predicted as not canceled, and the true positives represent the number of tasks correctly predicted as canceled. 
Moreover, accuracy and F1 score measure the overall correctness of the model’s predictions, where accuracy is the percentage of times the model correctly predicted the workers' behavior, and the F1 score is the harmonic mean of the precision and recall metrics. Precision is the proportion of tasks correctly predicted as canceled to the total number of tasks predicted as canceled. On the other hand, recall is the proportion of tasks predicted as canceled to the total number of canceled tasks.

Besides evaluating the behavioral models' performance, the effectiveness of the attack in evading detection is assessed by finding the average number of poisoning points detected as outliers by the autoencoder-based outlier detector  \cite{AE_anomaly}. Furthermore, the impact of the attack on worker selection is evaluated using the following metrics, as explained in section \ref{sec:behavioral selection}: the probability of task cancellation by workers derived from the behavioral model, the total payments made to victim workers,  and the QoS achieved for the task.

\subsection{Impact of $\alpha$ on the effectiveness of the attack}
\label{sec:impact of alpha}

The main goal of the attack is to increase the FPR of the workers' behavioral models, causing them to mislabel more samples from class 0 as 1. To determine the optimal value of alpha, a unique PGAN model was trained for each of the 86 workers in the dataset, across a range of $\alpha \in [0,1]$ with increments of 0.1. These models were used to generate poisoning points, which were then injected into each worker's training dataset by replacing features of the target class with the generated data. The resulting poisoned datasets were used to train workers’ behavioral models, and the effectiveness of the attack was evaluated on each worker's test dataset.

\begin{figure}
    \centering
    \includegraphics[width=0.75\linewidth]{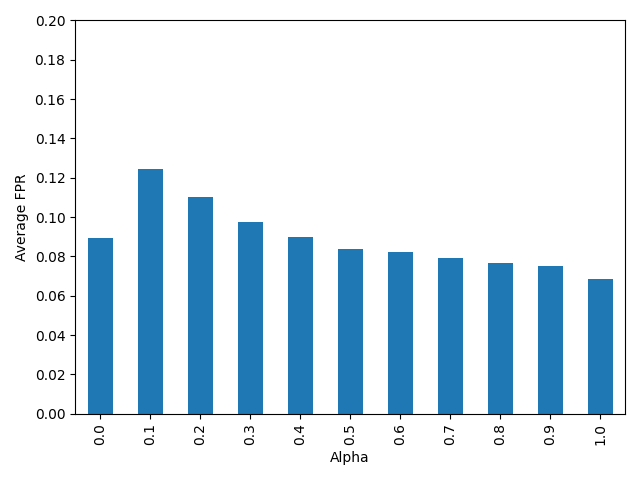}
    \caption{Average FPR of the workers' poisoned models for varying values of alpha}
    \label{fig:fpr_vary_alpha}
\end{figure}

\begin{figure}
    \centering
    \includegraphics[width=0.75\linewidth]{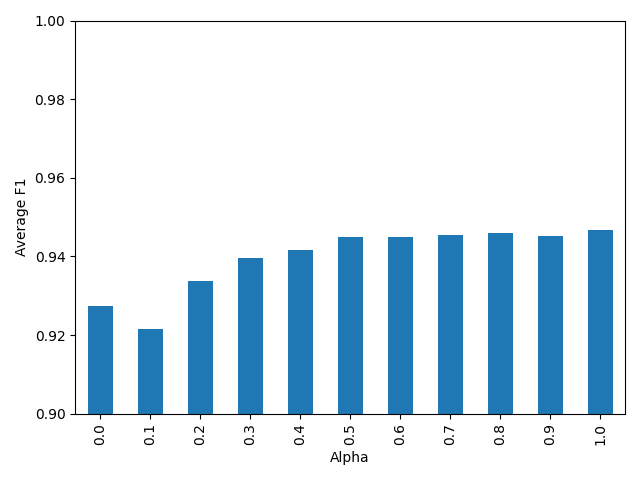}
    \caption{Average F1 scores of the workers' poisoned models for varying values of alpha}
    \label{fig:f1_vary_alpha}
\end{figure}

As illustrated in Figure \ref{fig:fpr_vary_alpha}, the FPR of the behavioral models varies significantly with $\alpha$. For instance, at $\alpha=1$, the injection of the generated points to the workers' training datasets results in the lowest FPR. This is because the generated points closely resemble the target class. However, lower alpha values resulted in higher FPRs, with the peak value of 0.12 achieved at $\alpha=0.1$. Therefore, this proves that this is the optimal value for $\alpha$, as it most effectively exploits the decision boundary of the behavioral models, thereby maximizing the success of the attack. Figure  \ref{fig:f1_vary_alpha} further illustrates the impact of the attack on the performance of the behavioral models by presenting the F1 scores. It can be observed that, at $\alpha=0.1$, the F1 reaches its lowest point, with a value of 0.92, which further proofs the effectiveness of the attack at this $\alpha$ value.

\begin{figure}[h]
    \centering
    \begin{subfigure}[b]{0.49\linewidth}
        \centering
        \includegraphics[width=\linewidth]{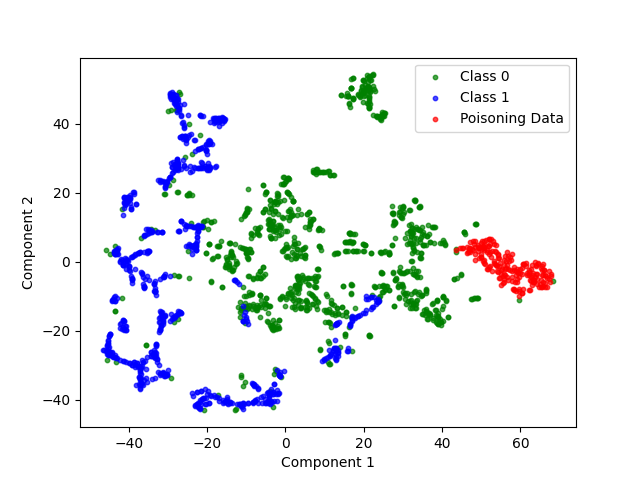}
        \caption{t-SNE plot with $\alpha = 0$}
        \label{fig:tsne0}
    \end{subfigure}
    \hfill
    \begin{subfigure}[b]{0.49\linewidth}
        \centering
        \includegraphics[width=\linewidth]{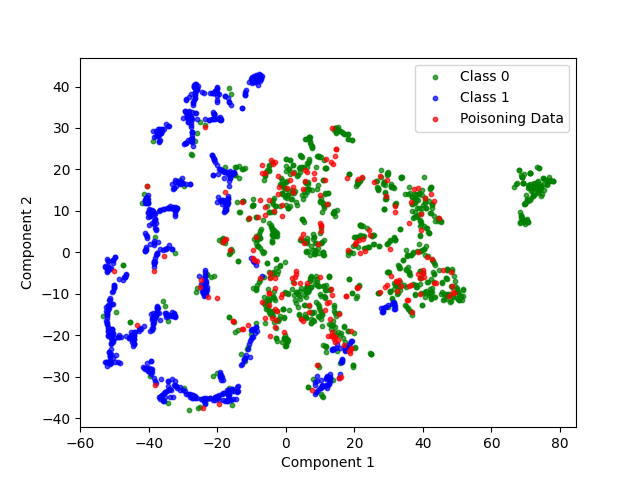}
        \caption{t-SNE plot with $\alpha = 0.1$}
        \label{fig:tsne0.1}
    \end{subfigure}
    
    \vspace{1em} % Adjust vertical space between rows

    \begin{subfigure}[b]{0.49\linewidth}
        \centering
        \includegraphics[width=\linewidth]{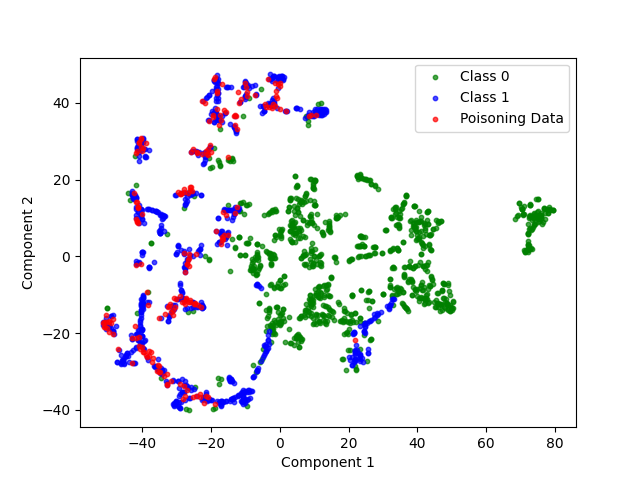}
        \caption{t-SNE plot with $\alpha = 1$}
        \label{fig:tsne1}
    \end{subfigure}
    
    \caption{T-sne plots showing the distribution of poisoning points for different $\alpha$ values: (a) $\alpha = 0$, (b) $\alpha = 0.1$, and (c) $\alpha = 1$.}
    \label{fig:tsne_combined}
\end{figure}

The value of alpha also plays a crucial role in shaping the distribution of the generated points. To further demonstrate its effect, the t-Distributed Stochastic Neighbor Embedding (t-SNE) dimensionality reduction technique was applied to one of the workers’ poisoned datasets. As shown in Figure \ref{fig:tsne0}, when $\alpha=0$, the generated poisoning points are positioned further from the target class (class 1) and exhibit more similarity to the other class (class 0). This is because, at this $\alpha$ value, no importance is given to generating points that are similar to the target class to evade detection. Instead, the model prioritizes generating points that compromise the model's performance. Figure \ref{fig:tsne0.1} illustrates the distribution of the original and poisoning points generated using $\alpha=0.1$. It can be observed that the poisoning points are spread between class 0 and class 1, reflecting the model's goal is to generate points that both compromise the model's performance and partially resemble the target class to evade detection. Additionally, the points generated with $\alpha =1$, as shown in Figure \ref{fig:tsne0.1}, closely resemble the target class. In this case, the PGAN behaves as a CGAN to generate points similar to class 1 without compromising the model's performance.   

To further show the impact of alpha across different behavioral patterns, t-SNE was applied to each worker’s poisoned dataset. After reducing the data to two dimensions, the centroid of class 1 data was evaluated. Subsequently, the Euclidean distance was calculated between each class 1 point and its centroid. Finally, a threshold of 5\% is applied to these distances to identify the most dissimilar points to the target class.

\begin{figure}
    \centering
    \includegraphics[width=0.70\linewidth]{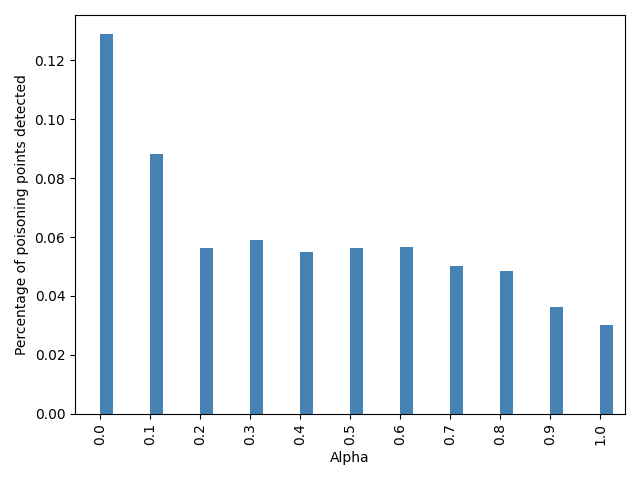}
    \caption{Fraction of poisoning points detected averaged for all workers' models}
    \label{fig:ppremoved}
\end{figure}

Figure \ref{fig:ppremoved} shows the percentage of the poisoning points deviating from the target class distribution, where each data point is the outcome of averaging the results for all workers. As shown in the figure, the highest fraction is observed at  $\alpha =0$. This is because the generator prioritizes increasing the error of the target classifier without any constraints related to the detectability of the generated poisoning points. As the value of $\alpha$ increases, less weight is given to reducing the error of the classifier, and more weight is given to increasing the error of the discriminator. Therefore, the percentage of poisoning points detected starts decreasing until it reaches its lowest value at $\alpha=1$. At this value, the generator's goal is to generate points that increase the error of the discriminator only. Consequently, the generated points become more similar to the points of the target class and the PGAN effectively behaves like a CGAN.

\subsection{Impact of varying poisoning percentages on the model performance}
\label{sec:model_performance_results}

\begin{figure}[h]
    \centering
    \includegraphics[width=0.6\linewidth]{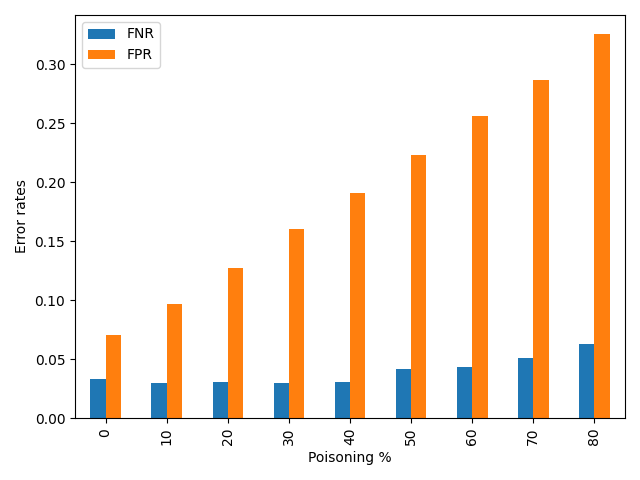}
    \caption{Models' average error rates for varying poisoning percentages}
    \label{fig:errorrateVarypp}
\end{figure}

Next, the workers' training datasets were poisoned with varying percentages of the target class data. The poisoning points were generated by GAN models trained using $\alpha = 0.1$ and $\lambda = 0.8$. Figure \ref{fig:errorrateVarypp} shows the average FPR and FNR of the workers' models for varying poisoning percentages. A clear increase in the FPR is observed, starting at 7\% with no poisoning points injected and rising to 33\% at 80\% poisoning. This highlights the effectiveness of the poisoning strategy in degrading the model’s performance, resulting in the misclassification of inputs belonging to class 0 as 1. Additionally, the FNR remains low and relatively unaffected, ensuring the attack remains focused on increasing the FPR.

\begin{figure}[h]
    \centering
    \includegraphics[width=0.6\linewidth]{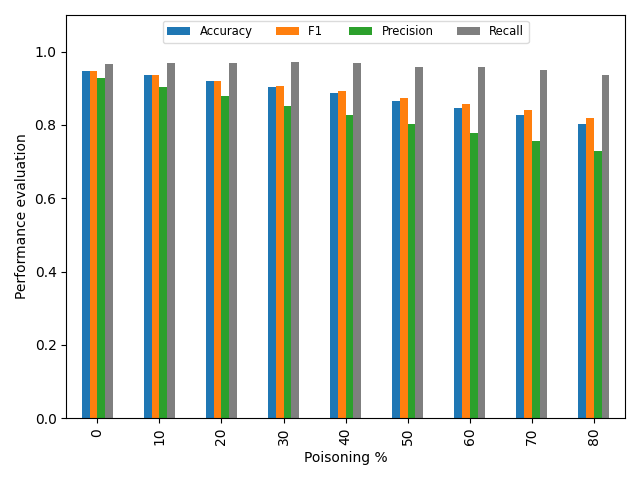}
    \caption{Models performance evaluation averaged for all workers for varying poisoning percentages}
    \label{fig:modelEval_varypp}
\end{figure}

Additionally, Figure \ref{fig:modelEval_varypp} shows that accuracy, F1 score, and precision decrease significantly as a result of the attack. Compared to the model trained on clean data, accuracy precision and F1 drop by 15\%, 20\%, and 13\%, respectively, at 80\% poisoning. However, recall remains relatively stable, decreasing by only 3\% at 80\% poisoning. This indicates that while the attack significantly affects the model’s overall predictive performance, its ability to identify true positives is only slightly impacted. This outcome aligns with the objective of the attack, which focuses on flipping negative class labels to positive while minimizing interference with the predictions of the positive class.

\begin{figure}[h]
    \centering
    \includegraphics[width=0.6\linewidth]{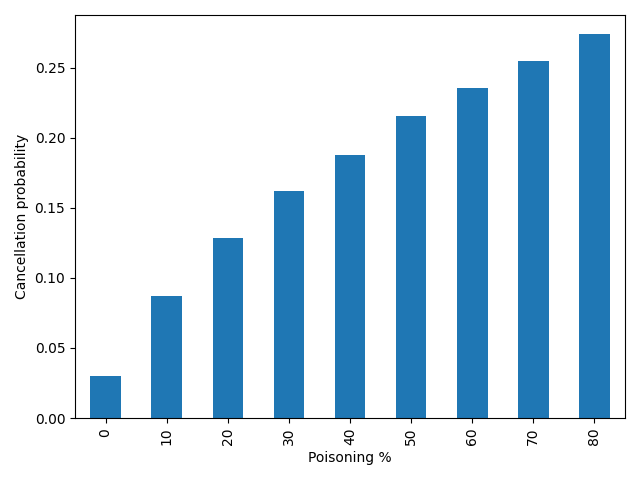}
    \caption{Task cancellation probabilities averaged for all workers}
    \label{fig:class0pc}
\end{figure}

Besides looking at the class predictions, it is also important to consider the impact of the attack on the task cancellation probabilities.  Figure \ref{fig:class0pc} shows the cancellation probabilities averaged for all workers for varying poisoning percentages. As illustrated, these probabilities increase as the poisoning percentage increases. This supports the attack's objective of reducing the victim workers' chances of selection. In addition, this also shows that even if the predicted labels don’t change, the attack can still affect the selection process, as the algorithm relies on probabilities rather than solely on labels.

\subsection{Comparison with Benchmark: Detectability and Model Performance}
\label{sec:benchmark_results}

In this section, the proposed attack is compared to \cite{aml_enviornmentmonitoring}, which utilized the label-flipping attack method. Label flipping is a commonly used technique to poison AI models during the training phase. The label-flipping attack was implemented by randomly selecting a fraction of data points from class 0 and flipping their labels to 1. The proposed attack is also compared against the feature manipulation attack, where noise is added to randomly selected data points from class 0 to shift them closer to the distribution of class 1. This attack does not alter the original class labels \cite{cleanlabel}.

The performance of the attacks is compared based on two aspects: their detectability by an autoencoder-based outlier detector and their impact on the models' performance. A consistent approach in training the outlier detector models was adopted for all attack methods to ensure a fair comparison of their detectability. The training of all autoencoder models was conducted for the same number of epochs and using the same architecture. The encoder network includes a fully connected dense layer with a ReLU activation function, while the decoder consists of another fully connected layer followed by a Sigmoid activation function. Each autoencoder model was trained for a total of 50 epochs. Additionally, to assess the impact of the attacks on the performance of the workers’ poisoned models, all classifiers were trained for the same number of epochs and using the same model architecture. The number of epochs used in the simulations is 2000.

\begin{figure}[htbp]
  \centering
  \begin{subfigure}[b]{0.45\textwidth}
    \centering
    \includegraphics[width=\textwidth]{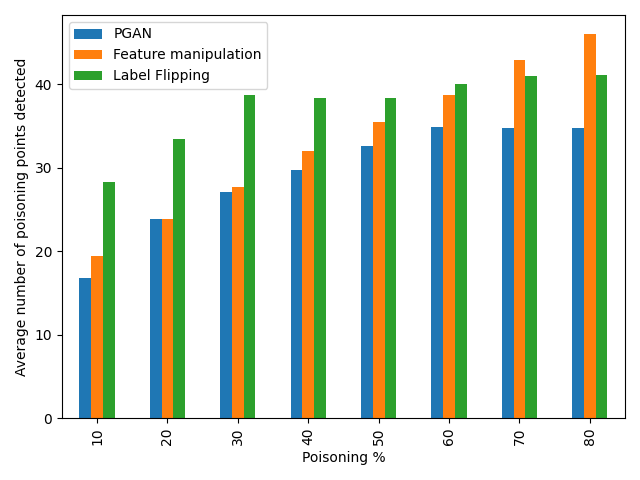}
    \caption{Average number of poisoning points detected with 5\% threshold.}
    \label{fig:PointsDetected_0.05}
  \end{subfigure}
  \hfill
  \begin{subfigure}[b]{0.45\textwidth}
    \centering
    \includegraphics[width=\textwidth]{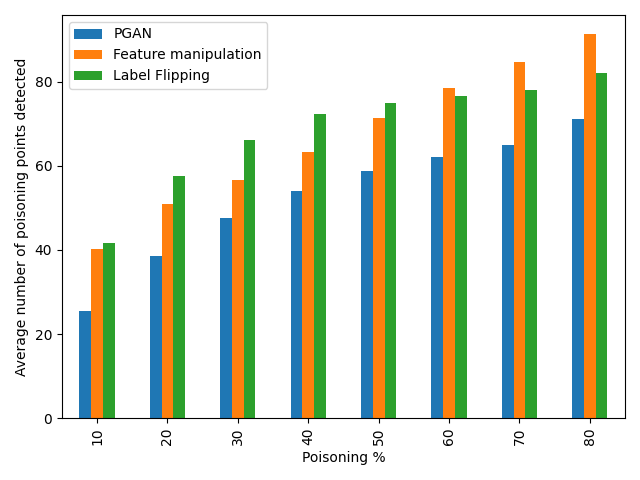}
    \caption{Average number of poisoning points detected with 10\% threshold.}
    \label{fig:PointsDetected_0.1}
  \end{subfigure}
  \caption{Average number of poisoning points detected by an autoencoder}
  \label{fig:PointsDetected}
\end{figure}

Figures \ref{fig:PointsDetected_0.05} and \ref{fig:PointsDetected_0.1} show the average number of poisoning points detected by an autoencoder-based outlier detector using 5\% and 10\% thresholds, respectively. This aligns with commonly used threshold values in the literature for anomaly detection, as in \cite{AEoutlierdetector}.  As illustrated, the proposed PGAN-based attack yields a lower average number of poisoning points detected at both threshold levels, which demonstrates its effectiveness in generating subtle and less detectable poisoning points, thus making it a more stealthy attack.

\begin{figure}[h]
    \centering
    \includegraphics[width=0.7\linewidth]{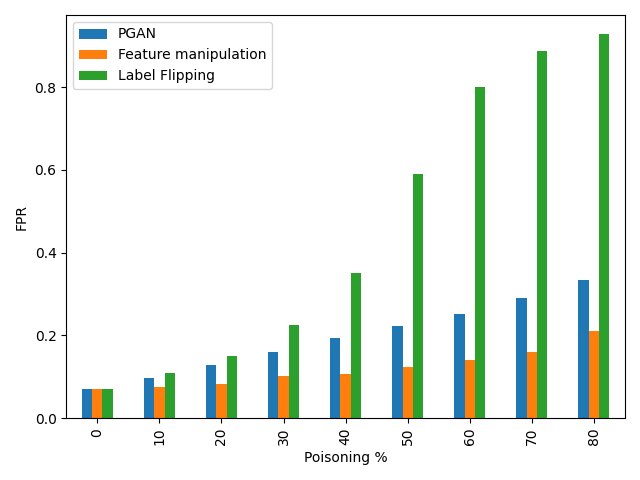}
    \caption{FPR of the proposed method and the benchmarks for varying poisoning percentages}
    \label{fig:fprbenchmark}
\end{figure}

\begin{figure}[h]
    \centering
    \includegraphics[width=0.7\linewidth]{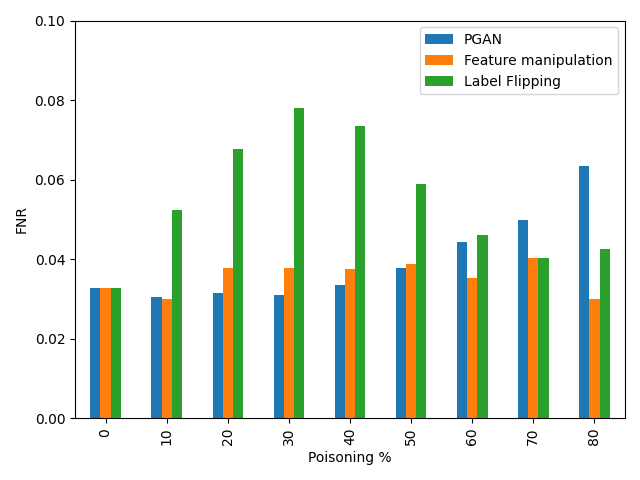}
    \caption{FNR of the proposed method and the benchmarks for varying poisoning percentages}
    \label{fig:fnrbenchmark}
\end{figure}

Moreover, Figures \ref{fig:fprbenchmark} and \ref{fig:fnrbenchmark} show the FPR and FNR of the poisoned behavioral models using the proposed, label flipping, and feature manipulation attacks, respectively. As illustrated in the figures, the proposed attack demonstrates superior performance compared to both benchmark approaches. Firstly, it achieves higher FPRs than the feature manipulation attack. In addition, although the label flipping attack achieves higher FPRs compared to the PGAN-based attack, the proposed method maintains the FNRs closer to the original value obtained without any poisoning, thus achieving its intended goal more effectively.

\subsection{Impact of the attack on the selection}
\label{sec:selection_results}
 
The experimental results presented in this section assess the impact of the attack on worker selection across a range of tasks, thereby demonstrating that a one-time poisoning of victim workers' models can lead to significant long-term effects. To assess the effectiveness of the attack, the total payment made to the victim workers after completing 100 tasks was evaluated, thereby effectively reflecting the diverse nature of tasks from the real-world. Additionally, 20\% of the total workers in the dataset were randomly selected as the victims.

\begin{figure}[h]
    \centering
    \includegraphics[width=0.7\linewidth]{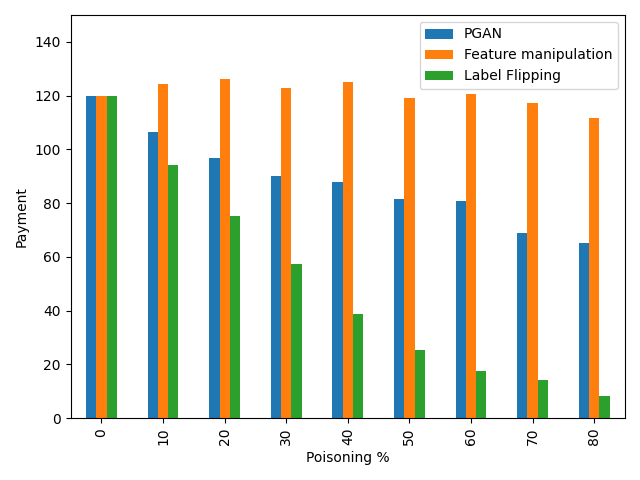}
    \caption{Payment received by the victim workers for 100 tasks, averaged across all workers}
    \label{fig:tot_payment_victim}
\end{figure}

The results were then compared with the payment values obtained under the label flipping and feature manipulation attacks.  
Figure \ref{fig:tot_payment_victim} presents the total payment received by the victim workers for 100 tasks, averaged across all workers. As illustrated, in a normal scenario with no poisoning points injected into the training dataset, the average payment received by the victim workers is 120. As the poisoning percentage increases, the proposed attack significantly reduces the total payment received due to being selected for fewer tasks. In contrast, the feature manipulation attack does not achieve the same effect. Additionally, although the PGAN-based attack results in higher payment values than the label-flipping attack, it is less detectable because of its more gradual reduction in payment as the poisoning percentage increases. The payment reduction can reach up to 47\% at 80\% poisoning, using the proposed PGAN-based attack. On the other hand, the reduction in payment achieved by label flipping at the same poisoning percentage can reach up to 93\%. Consequently, the proposed attack is considered to be more successful, as it reduces the payment of victim workers while evading detection more effectively. 

%In contrast, for the non-victim workers, the total payment increases as the poisoning percentage increases, as shown in Figure \ref{fig:tot_payment_nonvictim}. This is expected since reducing the likelihood of selecting victim workers increases the probability of selecting non-victim workers, thus increasing their total payment.

\begin{figure}[h]
    \centering
    \includegraphics[width=0.7\linewidth]{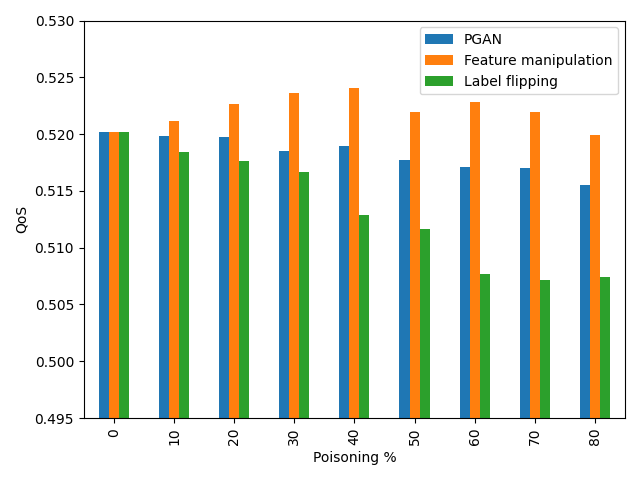}
    \caption{$QoS_g$ averaged for all tasks}
    \label{fig: qos_g}
\end{figure}

Figure \ref{fig: qos_g} shows $QoS_g$ averaged for all tasks for varying poisoning percentages. It can be seen that although more poisoning points are introduced, the PGAN-based attack maintains relatively stable QoS values. On the other hand, the decrease in the QoS values obtained with the label-flipping attack is more significant. This proves the effectiveness of the proposed attack in bypassing other layers of detection.

\subsection{Discussion}

\subsubsection{Summary of key findings}
Based on the results presented in sections~\ref{sec:impact of alpha}-\ref{sec:selection_results}, it was shown that the proposed PGAN-based attack demonstrates superior performance compared to existing benchmark methods in multiple aspects. Firstly, it generates more subtle and less detectable poisoning points by an autoencoder outlier detector. Secondly, unlike label flipping attack, the proposed method is able to meet the objective of the attack more effectively by predicting that the victim workers will cancel the task when they actually are more likely to accept it. This is done while ensuring that the poisoned model does not erroneously predict task acceptance when the victim worker is, in fact, more likely to cancel. In other words, the proposed method effectively increases the FPR while preserving the FNR.

Thirdly, the PGAN-based attack successfully reduces the total payment received by victim workers. However, unlike the benchmark label flipping attack, which results in a sharp and easily detectable decline in payment, the proposed method introduces this reduction more gradually.  This further proves the superior performance of the proposed attack in terms of stealthiness, since a gradual reduction in payment makes it less likely for system administrators to suspect that an insider attack took place and can be attributed to normal fluctuations in worker performance. For instance, consider a scenario where a malicious insider poisons the victim workers’ models with 40\% poisoning percentage. As shown in Figure \ref{fig:tot_payment_victim}, using label flipping attack, the average payment received by the victim workers after completing 100 tasks drops sharply to 40, compared to 120 in a normal scenario. In contrast, the PGAN-based attack reduces the payment to 90, a decline that is less drastic and is, therefore, less likely to raise suspicion from the victim workers or the management platform.

\subsubsection{Deployment challenges and implications of the proposed attack}

One of the key challenges in executing the proposed attack lies in tuning PGAN parameters, such as $\alpha$, to ensure the effectiveness and stealthiness of the attack. Moreover, the proposed attack in this paper has broader implications on the MCS system, beyond just the reduction in the revenue of the victim workers. It also poses a serious threat to the platform as it can, over time, lead to losing trust in the system's reliability and trustworthiness. In fact, real-world MCS systems already face challenges in earning and maintaining workers' trust, as workers may suspect that malicious insiders with access to sensitive resources are manipulating the system’s decisions \cite{AMLiot}. Prior studies further highlighted this concern. For instance, in \cite{UberBias} and \cite{cbsnews_algorithmic_wage_discrimination}, workers in one of the ride-sharing platforms have reported experiences suggesting possible bias in the selection algorithm. For instance, some drivers shared that after completing nearly all the trips needed to earn a \$100 bonus, they faced an unusually long wait for the final ride, despite being in a busy area. As a result, this lead them to question the fairness of the system and caused them to start losing trust in the platform.

\subsubsection{Defense mechanisms}

To mitigate potential GAN-based attacks, several defense mechanisms can be adopted. Some of them serve as generic countermeasures, including conventional techniques like cryptographic methods and differential privacy approaches, which help protect sensitive data within the system. By implementing robust privacy-preserving measures, such as those used in federated learning systems, the risks of GAN-based attacks can be mitigated, thus improving the platform’s trustworthiness \cite{FLpaper}. However, these approaches come with trade-offs, which include increased computational cost and degradation in the performance of the machine learning models. The platform can also utilize methods that enhance the robustness of the system, such as ensemble learning and adversarial training. In ensemble learning, multiple models are trained, and their aggregated predictions are utilized after deployment, thus reducing the impact of poisoning attacks. In addition, in adversarial training, GAN-generated data are incorporated during the training stage to enhance the model's robustness against adversarial threats \cite{gan_attacks_survey}.

\section{Conclusion}

In this paper, a novel adversarial attack on behavioral-based MCS worker selection is proposed, where insider adversaries inject stealthy poisoning points into victim workers' datasets to reduce their revenue. By leveraging GANs, the attack targets vulnerable regions in the feature space to degrade model performance while ensuring the poisoning points remain undetected by outlier detectors. Simulation results using a real-life dataset demonstrate the effectiveness of the attack. First, the impact of $\alpha$ on the detection rate of poisoning points and model performance was examined. Second, the impact of varying poisoning percentages on the model performance was assessed. Finally, the attack’s effectiveness in reducing victim workers' payment was evaluated for different poisoning percentages.

\bibliography{ref.bib}

 \end{document}